\documentclass[twocolumn,prl,amsmath,amssymb]{revtex4}
\usepackage{graphicx}% Include figure files
\usepackage{dcolumn}% Align table columns on decimal point
\usepackage{bm}% bold math

\begin{document}

\preprint{APS/123-QED}

\title{Cold collision shift cancelation and inelastic scattering in a Yb optical lattice clock}

\author{A.~D.~Ludlow, N.~D.~Lemke, J.~A.~Sherman, and C.~W.~Oates}
\affiliation{National Institute of Standards and Technology, 325
Broadway, Boulder, CO  80305, USA}
\email{ludlow@boulder.nist.gov}
\author{G.~Qu\'{e}m\'{e}ner, J.~von Stecher, and A.~M.~Rey}
\affiliation{JILA, National Institute of Standards and Technology and University of Colorado, Boulder, CO 80309-0440}

\begin{abstract}
Recently, $p$-wave cold collisions were shown to dominate the density-dependent shift of the clock transition frequency in a $^{171}$Yb optical lattice clock.  Here we demonstrate that by operating such a system at the proper excitation fraction, the cold collision shift is canceled below the $5 \times 10^{-18}$ fractional frequency level. We report inelastic two-body loss rates for $^3\!P_0\,$-$^3\!P_0$ and $^1\!S_0\,$-$^3\!P_0$ scattering.  We also measure interaction shifts in an unpolarized atomic sample.  Collision measurements for this spin-1/2 $^{171}$Yb system are relevant for high performance optical clocks as well as strongly-interacting systems for quantum information and quantum simulation applications.
\end{abstract}
\maketitle

Large ensembles of ultracold atoms offer atomic clocks a measurement of the atomic state with high signal-to-noise ratio.  For clocks utilizing optical transitions, this has the potential to yield time and frequency measurements with new levels of precision and speed (e.g.\ \cite {JiangLudlow2011,TakamotoTakano2011}).  However, large ensembles of cold atoms can lead to high number density and thus significant interatomic interactions.  These interactions can perturb the clock transition frequency, compromising the accuracy of the atomic standard.  For example, in cesium fountain primary standards, cold collision shifts can become significant \cite{GibbleChu1993,LeoJulienne2001}, influencing clock operation (e.g.\ \cite{DosSantosMarion2002,HeavnerJefferts2005,SzymaniecChalupczak2007}).

Optical lattice clocks, which probe the ultra-narrow $^1\!S_0\,$-$^3\!P_0$ transition in two-valence-electron atoms held in an optical potential, are also susceptible to cold collisions.  Non-negligible collision effects have been observed in lattice clocks using nuclear-spin-polarized, fermionic samples of $^{87}$Sr \cite{LudlowZelevinsky2008,CampbellBoyd2009} and $^{171}$Yb \cite{LemkeLudlow2009}.  Measurement and control of these collisions therefore play a key role in the continued development of these standards.  At the same time, the control of these interactions are an integral part of proposals for quantum information \cite{HayesJulienne2007,GorshkovRey2009} and quantum simulation of solid-state-analog Hamiltonians \cite{GorshkovHermele2010,CazalillaHo2009,FossFeigHermele2010}.  For $^{87}$Sr, the primary interaction giving the cold collision shift was identified as an $s$-wave interaction between atoms in non-identical superpositions of the clock states \cite{CampbellBoyd2009,ReyGorshkov2009,Gibble2009}.  It was recently observed that strong interactions can isolate and suppress the shift \cite{ReyGorshkov2009,SwallowsBishof2011}.

For $^{171}$Yb, it was shown that a $p$-wave interaction between atoms in the $^1\!S_0$ and $^3\!P_0$ electronic states was the dominant mechanism responsible for the cold collision shift \cite{LemkeLudlow2011}.  Unlike the $s$-wave case, such an interaction is less sensitive to small particle distinguishability between the ultracold fermions.  Consequently the observed shift exhibits roughly a linear dependence on the $^3\!P_0$ excitation fraction (Fig.~1(a)).  Moreover, in the weakly interacting regime of the 1-D lattice, the cold collision shift crossed zero at a mean excitation fraction close to 0.5.  In this work, we exploit this zero-crossing to demonstrate cancelation of the $^{171}$Yb cold collision shift in a 1-D optical lattice below the $5\times 10^{-18}$ fractional frequency level. The cancelation is enabled by the fact that near 50$\%$ excitation, population of the two electronic states of one atom induce equal collision shifts on the opposite states of another atom, leaving zero net shift for the clock transition \cite{Footnote1}.  The collision shift cancelation can thus be likened to the Stark shift cancelation which optical lattice clocks exploit by operating at the ``magic'' wavelength \cite{KatoriTakamoto2003,YeKimble2008}.  Our uncertainty in this cancelation reaches below the smallest total uncertainty levels reported to date for any type of atomic clock, making it a powerful and pragmatic technique for mitigating the collisional shift in a lattice clock.

The $^{171}$Yb ($I=1/2$) optical lattice clock is described in detail elsewhere \cite{LemkeLudlow2011,LemkeLudlow2009}.  After dual laser-cooling stages, atoms are loaded into a 1-D optical lattice using a laser at $\lambda_{\text{magic}}\simeq759$ nm. Atomic population, initially split between the two ground magnetic substates, is optically pumped to a single spin state using $\sigma$-polarized light resonant with the $^1\!S_0$ ($I=1/2$) - $^3P_1$ ($I=3/2$) transition in the presence of a bias magnetic field (5~G).  We estimate the average atomic density can reach $\rho_0 \approx 3 \times 10^{11}$ atoms/cm$^3$, with an atomic temperature of $\sim10$ $\mu$K.  The atoms are then spectroscopically probed on the $^1\!S_0\,$-$^3\!P_0$ transition using Ramsey spectroscopy.  The Ramsey pulse times used here are 5-10 ms, while the dark time is 80-150 ms.  A cavity-stabilized probe laser is actively stabilized to the center Ramsey fringe using an acousto-optic modulator.  The cold collision shift is the measured frequency difference between interleaved atomic samples of high and low density.

\begin{figure}[t]
\resizebox{8.5cm}{!}{
\includegraphics[angle=0]{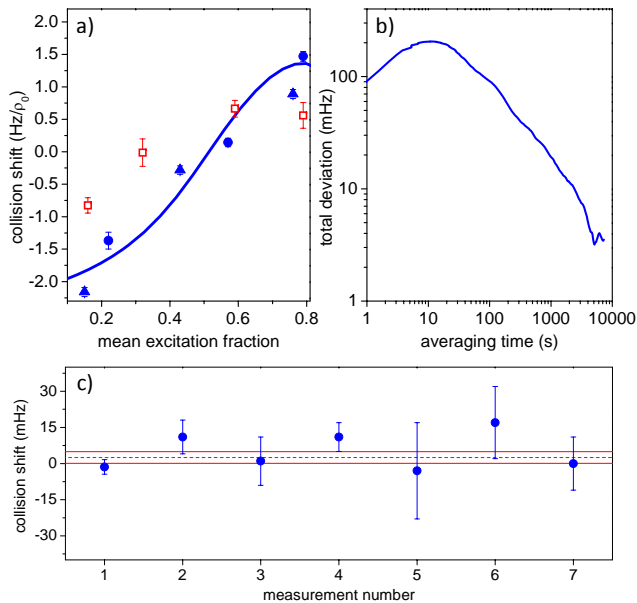}}
\caption{\label{Fig1}(color online) (a) Cold collision shift as a function of excitation fraction,  for a polarized sample in a 1-D optical lattice oriented vertically (filled triangles) or horizontally (filled circles). Open squares are with an unpolarized atomic sample.  (b) Cold collision shift cancelation:  the measurement uncertainty for this cancelation is shown, given by the total deviation (similar to two-sample Allan deviation \cite{GreenhallHowe2011}).  (c) Measurement of the residual cold collision shift when operating at $51\%$ excitation and an atomic density of $\rho_1$.  The red dashed line indicates the weighted mean of the seven measurements, +2.5 mHz, while the solid lines gives the one-$\sigma$ error bars of $\pm 2.4$ mHz.}
\end{figure}

Measurement of the collision shift in a 1-D lattice is shown as filled points in Fig.~1(a), as a function of excitation fraction (during the Ramsey dark time).  The theoretical calculation of the shift using a $p$-wave model is also shown (solid line). In the model, the dominant interaction, $V_{eg}$, is between $^1\!S_0$ and $^3\!P_0$ atoms, and a weaker interaction between two excited state atoms, $V_{ee}=0.1 \times V_{eg}$, is also included  \cite{LemkeLudlow2011}.  The shift has a zero value near 50$\%$ excitation, and this zero-crossing is the focus of our attention here.  To determine the precise excitation corresponding to zero shift, we made real-time measurements of the excitation fraction by turning off the second Ramsey pulse and measuring the ground and excited atomic populations after the dark time. These measurements were interspersed between cycles of usual Ramsey spectroscopy. We then slightly adjusted the probe laser power to keep the measured excitation fraction constant during each shift measurement.

The precision of the zero-crossing measurement is given by the clock instability during interleaved measurements of high and low atomic density.  Our measurement stability benefits from recent improvements to the cavity-stabilized laser used to probe the clock transition \cite{JiangLudlow2011}.  Fig.~1(b) shows the precision of one such measurement, where we observed a collision shift of -1.4 (3) mHz after 14500~s of averaging. Fig.~1(c) shows the result of seven similar measurements, taken sequentially over the course of several weeks and with different measurement durations.  For all but one measurement, the mean excitation fraction was controlled to $51\pm 0.3 \pm1.3$ $\%$, where the first uncertainty is the fluctuation in the measured excitation fraction and the second is the systematic uncertainty in the absolute excitation value.  (For data point number two, the excitation fraction was set to $1\%$ higher; we thus applied a $-5$ mHz correction to the measured collision shift, as determined from the slope of the curve in Fig.~1(a).)  The weighted mean of the seven measurements (red lines) is 2.5 (2.4) mHz (reduced $\chi^2=1.04$).  This corresponds to a fractional shift of 4.8 (4.6) $\times 10^{-18}$ of the transition frequency, and demonstrates the smallest measurement of a collision shift in a lattice clock.

In order to routinely implement this collision shift cancelation, we consider the robustness of this technique to relevant experimental conditions.  For an operational density of $\rho_1 \simeq 3 \times 10^{10}$ atoms/cm$^3$, a 1$\%$ change in the excitation fraction leads to a change in collision shift of $\sim 1 \times 10^{-17}$.  As described above, it is straightforward to keep the excitation fixed at or below the $1 \%$ level.  However, a complication arises from time-dependent excitation due to trap loss.  Particularly, we have observed inelastic two-body losses involving both $^3\!P_0$-$^3\!P_0$ and $^1\!S_0\,$-$^3\!P_0$ interactions.  With both $^1\!S_0$ ($g$) and $^3\!P_0$ ($e$) populations present, the number density rate equations are:
\begin{eqnarray*}
  \dot{n}_g(t)= -\Gamma_g n_g(t) - \beta_{eg} n_g(t)n_e(t) \\
  \dot{n}_e(t)= -\beta_{ee} n_e(t)^2 - \Gamma_e n_e(t) - \beta_{eg} n_g(t)n_e(t)
\end{eqnarray*}
For a single population ($n_g$ or $n_e$) $\beta_{eg}$ loss can be ignored.  Fig.~2(a) shows $^1\!S_0$ trap loss (black triangles, for a pure sample of $^1\!S_0$ atoms), with a one body loss fit yielding a trap lifetime of $1/\Gamma_g=$ 480 (20) ms.  Also shown is $^3\!P_0$ trap loss (blue circles, for a pure sample of $^3\!P_0$ atoms).  This population experiences notably stronger decay at high densities, and good fit requires both one-body ($\Gamma_e$) and two-body ($\beta_{ee}$) losses.  After integrating the losses over the spatial extent of a single lattice site, as well as across the distribution of occupied sites, we find $1/\Gamma_e=$ 520 (28) ms and $\beta_{ee} = 5 \times 10^{-11}$ cm$^3$/s, somewhat larger than the $^3\!P_0$-$^3\!P_0$ decay measured in $^{88}$Sr \cite{TraversoChakraborty2009,LisdatWinfred2009}. Because the absolute atomic density is difficult to calibrate, the uncertainty in $\beta_{ee}$ is estimated as 60$\%$.

\begin{figure}[t]
\resizebox{8.5cm}{!}{
\includegraphics[angle=0]{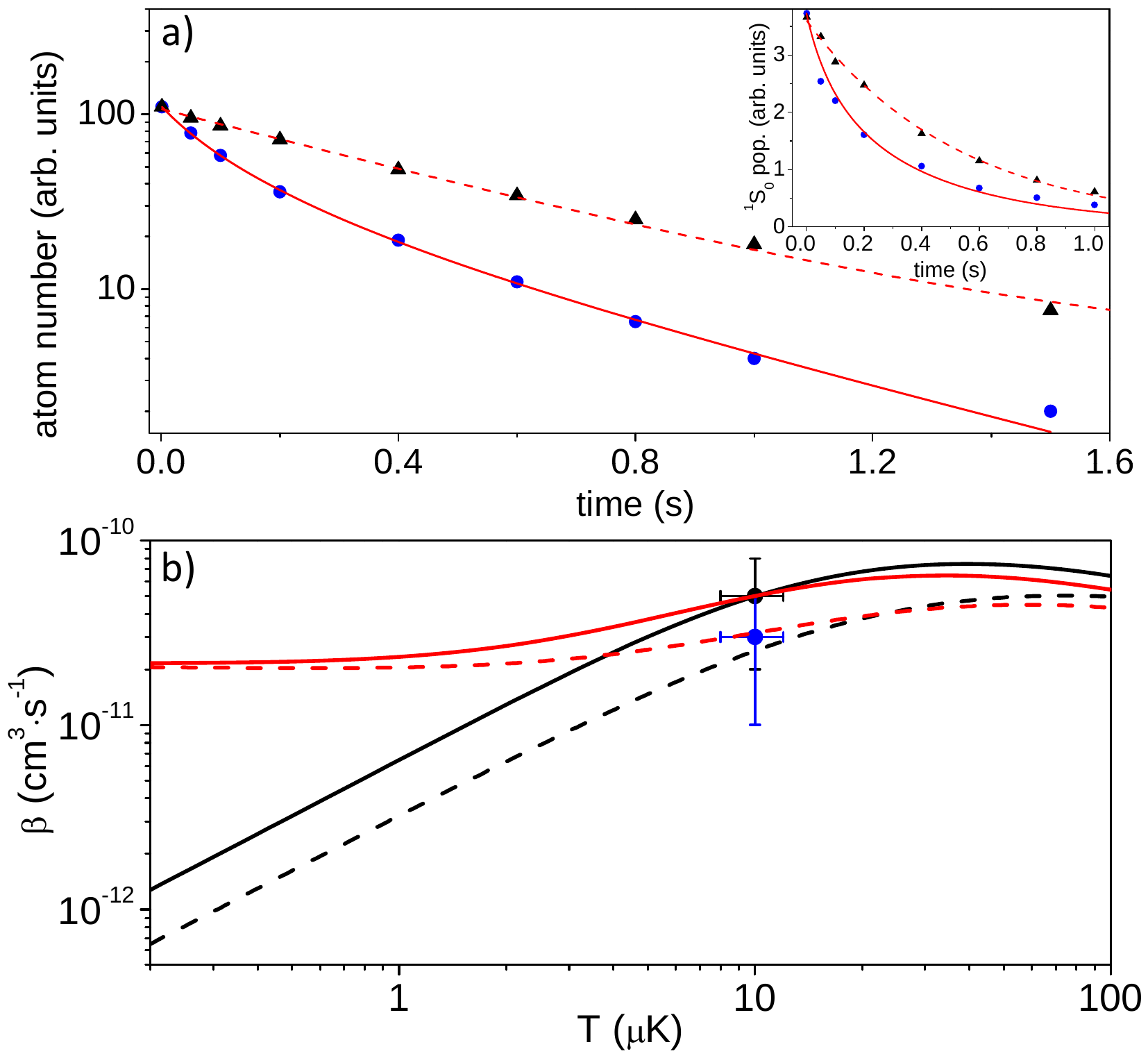}}
\caption{\label{Fig2}(color online) (a) $^1\!S_0$ trap loss (black triangles) with an exponential fit (dashed curve). $^3\!P_0$ trap loss (blue circles) with a one- and two-body loss fit (solid curve).  The inset shows $^1\!S_0$ loss for a pure $^1\!S_0$ sample (black triangles) and a mixed sample of $^1\!S_0$ and $^3\!P_0$ atoms (blue circles).  The former is fit with a simple exponential, the latter with $^3\!P_0\,$-$^3\!P_0$ and $^1\!S_0\,$-$^3\!P_0$ two body losses. (b) Two body loss rates as a function of temperature, calculated with $p_\text{ls}=1$ (dashed curves) and with $p_\text{ls}=0.8$ and $\delta=0.51 \, \pi$ (solid curves).  Black is for a nuclear-spin-polarized sample of $^3\!P_0$ atoms ($\beta_{ee}$), red an unpolarized sample of $^3\!P_0$ atoms ($[\beta_{ee}+\tilde{\beta}_{ee}]/2$).  Circles give experimental measurements, black for $\beta_{ee}$ and blue for $\beta_{eg}$.}
\end{figure}

Inelastic two-body loss for $^3\!P_0$ atoms can be considered with a single atom-atom channel, time-independent quantum  formalism similar to \cite{IdziaszekJulienne2010,IdziaszekQuemener2010}. Lacking an accurate Yb$_2$ potential, the short range physics is described here by a boundary condition at an interatomic separation $R_0=20$~a$_0$ (a$_0$ is the Bohr radius), represented by two parameters: (i) an accumulated phase shift, $ 0 \le \delta \le \pi$, due to the unknown short-range potential from $R=0$ to $R=R_0$, and (ii) a loss probability, $ 0 \le p_\text{ls} \le 1$ at $R=R_0$. The long range physics is accurately described by a van der Waals interaction, with $C_6 = 3886$ a.u.\ given by \cite{DzubaDerevianko2010}. The Schr\"odinger equation is numerically solved from $R_0$ to $R \to \infty$, and the cross section is computed as a function of the collision energy. The thermalized rate coefficients are found by averaging the cross sections with a Maxwell--Boltzmann distribution of collision energies for a given temperature.

With full loss probability ($p_\text{ls}=1$), $\beta_{ee}$ is shown in Fig.\ 2(b) (black dashed curve) as a function of temperature.  Here, the rates are universal, independent of $\delta$~\cite{IdziaszekJulienne2010,IdziaszekQuemener2010}. The interaction involves two identical fermions, and thus quantum statistics dictates that the interactions must be odd partial waves, notably $p$-wave at these ultracold temperatures. As two excited atoms approach each other, those which successfully tunnel through the $p$-wave barrier are assumed to inelastically scatter with unit probability, a reasonable assumption due to the large number of exit collision channels available to a pair of atoms with high internal energy. For an unpolarized sample of $^3\!P_0$ atoms (here with equal $m_I=\pm 1/2$ populations), interactions between distinguishable atoms also include an $s$-wave term, and the loss rate for distinguishable $^3\!P_0$ atoms is labeled $\tilde{\beta}_{ee}$. The total loss rate for the unpolarized sample, given by the average of $\beta_{ee}$ and $\tilde{\beta}_{ee}$, is shown in Fig.\ 2(b)  as a red dashed curve ($p_\text{ls}=1$). Experimentally, we observed identical loss rates within $10\%$ for a polarized and an unpolarized sample at 10 $\mu$K (Fig.\ 2(b), black circle).  Since full loss predicts unequal loss rates for polarized and unpolarized samples at 10 $\mu$K, $^3\!P_0$ atoms may not be lost with full unit probability at short range. Instead, using short range parameters of $p_\text{ls}=0.8$ and $\delta=0.51 \, \pi$, the loss rates are shown as solid curves in Fig.\ 2(b) for polarized (black) and unpolarized (red) cases. Here, we found better agreement with the experimental data, suggesting a deviation from the universal regime.

For a mixed population of $^1\!S_0$ and $^3\!P_0$ atoms, we observed additional loss through $\beta_{eg}$.
Fig.~2(a) inset highlights this by comparing $^1\!S_0$ trap loss for two different atomic samples: with (blue circles) and without (black triangles) the presence of $^3\!P_0$ atoms. The additional loss is particularly notable at short times where both $^1\!S_0$ and $^3\!P_0$ populations are large and is consistent with a $^1\!S_0\,$-$^3\!P_0$ two-body loss rate at the level of $\beta_{eg}=3 \times 10^{-11}$ cm$^3$/s.  The $^3\!P_0$ population is excited coherently from the $^1\!S_0$ state, but due to excitation inhomogeneity, the lossy collisions are between partially distinguishable atoms \cite{LemkeLudlow2011}.  The magnitude of the inelastic loss is noteworthy because for molecular states correlating to a $^1\!S_0$ and $^3\!P_0$ atom pair, only the  $^1\Sigma_g^+$ ground state (correlating to the $^1\!S_0$-$^1\!S_0$ state) lies at lower energy, and long-range coupling to this state is spin-forbidden.

In general, all of these loss processes lead to a time-dependent excitation fraction during the Ramsey dark time.  This, in turn, affects the balance between $^1\!S_0$ and $^3\!P_0$ collisionally induced energy shifts.  For simplicity, we can easily operate at a lower atom number density, reducing not only the collisionally-induced shifts, but also the two-body inelastic losses.  At the same time, the number of quantum absorbers remains well in excess of 1000, in order to accommodate a high signal-to-noise ratio.  At a density of $\rho_1$, the excitation fraction over a dark time of $T=150$~ms changes by only several percent.  We therefore take the time-averaged value to indicate the excitation fraction where the shift is canceled.

Any degree of imperfect polarization of the nuclear spin state introduces a host of other possible atomic interactions.  In $p$-wave, the $V_{eg}^{-}$ interaction, which shifts the singlet state, becomes allowed \cite{LemkeLudlow2011}.  Furthermore, distinguishability between different nuclear spin states allows $s$-wave interactions $U_{gg}$, $U_{ee}$, and $U_{eg}^{+}$.  In $^{171}$Yb, both the $s$- and $p$-wave $|gg\rangle$ interaction terms are small \cite{KitagawaEnomoto2008}.  However, the remaining interactions can alter the observed collision shift from the spin-polarized case.  For this reason, high polarization purity is important for optical lattice clocks.  Here, we benefit from the simple structure ($I=1/2$) of $^{171}$Yb and can readily optically pump to a single spin ground state with $99\%$ purity.  The polarization purity is directly measured by observing the absence of a clock excitation spectrum for the unpopulated spin state.

To quantify the effect of imperfect polarization, we measure the collision shift as a function of excitation for unpolarized atoms (open squares in Fig.~1(a), equal mixture of both spin states).  Since, during spectroscopy, a weak bias field (1 G) lifts the degeneracy of the $\pi$-transitions from the two spin states, here the excitation fraction is defined as that for the particular spin state being resonantly excited, not accounting for the other `spectator' spin state.  In general, the measured shifts are smaller than for the spin-polarized case, implying that competing interactions have the opposite sign as those in the polarized case.  Notably, the measured zero-crossing in the shift occurs at a lower excitation fraction (around $40\%$), leaving a net positive shift at $51\%$, where the shift is zero for the spin polarized case.  Based upon these measurements, we determine a $1\%$ polarization impurity will not affect the shift zero-crossing at or above the level measured in Fig.~1(c).

\begin{figure}[t]
\resizebox{8.5cm}{!}{
\includegraphics[angle=0]{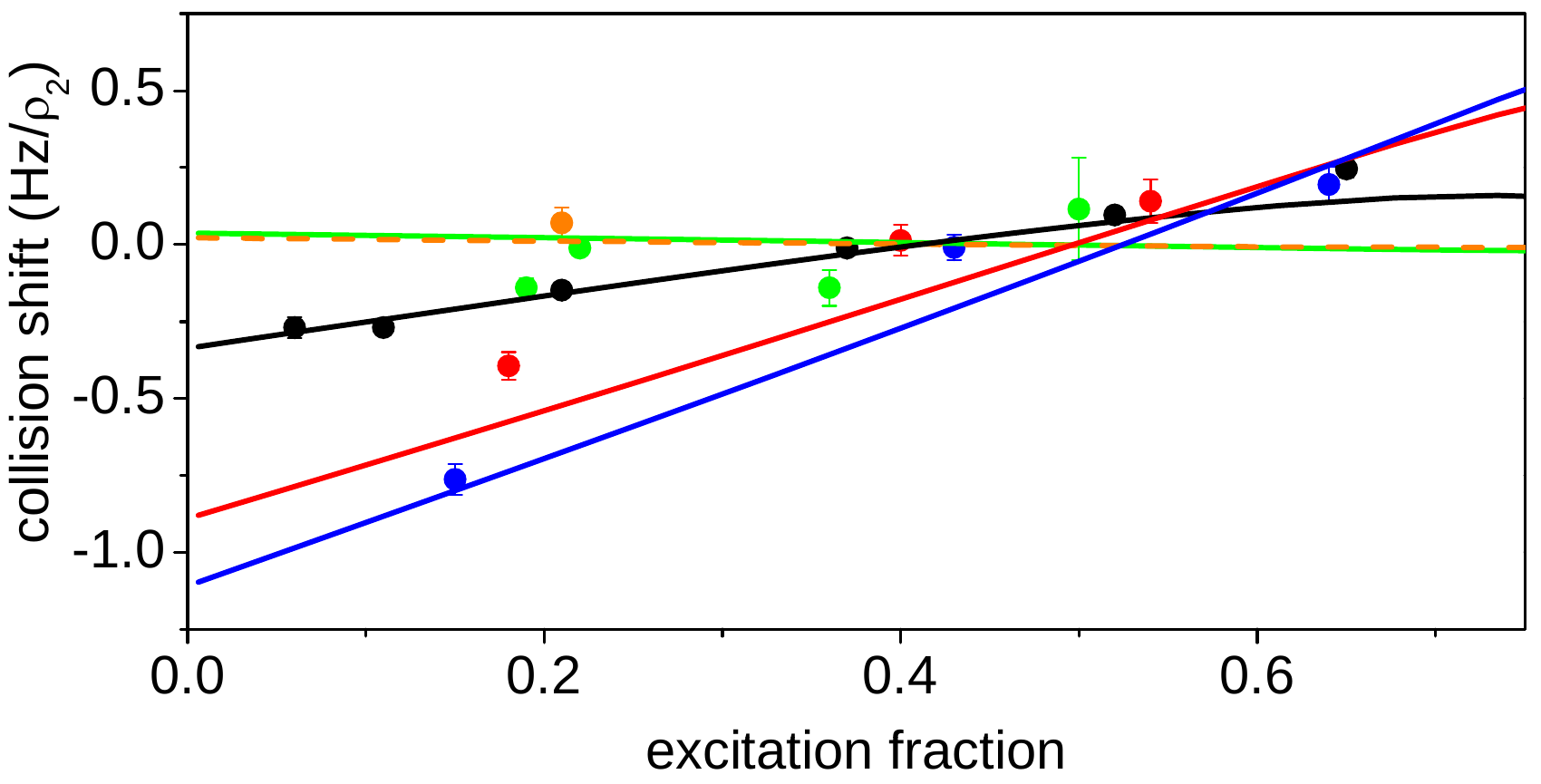}}
\caption{\label{Fig3}(color online) Cold collision shift as a function of excitation for a polarized sample in a 2-D lattice, for different Ramsey times.  $\rho_2$ is a typical 2-D lattice density, where $25\%$ of the atoms are doubly occupied in a lattice site with density of $4\times 10^{12}$ atoms/cm$^3$.  Circles correspond to experimental measurements, and solid curves are theoretical calculations \cite{LemkeLudlow2011}.  In both cases, blue is a Ramsey time of $T=10$ ms, red is 40 ms, black is 80 ms, green is 160 ms, and orange is 210 ms.}
\end{figure}

We have shown that the cold collision shift can be canceled at the $5\times 10^{-18}$ level in a 1-D optical lattice clock of $^{171}$Yb. Other implementations may be suitable for reducing the clock shift at or below this level.  In a 2-D optical lattice clock, strong interactions can lead to a decay of the collision shift \cite{LemkeLudlow2011,Gibble2010}.  In particular, longer Ramsey dark times lead to smaller shifts and can reduce the shift dependence on excitation fraction (Fig.~3), making it attractive for shift reduction. Care must be taken since, as we have observed both experimentally and theoretically, the interactions can also reduce Ramsey fringe contrast. Alternatively, the 2-D lattice system also exhibits a zero crossing in the shift versus excitation.  However, strong interactions can move the zero-crossing excitation with a nonlinear dependence on interaction strength (Fig.~3), and thus the weak interactions of the 1-D lattice may be easier to control.

The 3-D optical lattice continues to be an interesting choice \cite{KatoriTakamoto2003}, where it is straightforward to achieve high atom number with an average of $\ll1$ atom per lattice site.  The small fraction of lattice sites with double occupancy will exhibit very strong interactions. Doubly-occupied sites could also be eliminated using photoassociation losses \cite{AkatsukaTakamoto2010} or directly via the two-body losses observed here. Furthermore, the 3-D optical lattice system may exhibit kinematic suppression of the atomic interactions which yield the shifts \cite{GunterStoferle2005}.  Indeed, these rich atomic systems will undoubtedly continue to offer many interesting phenomena in 1-, 2-, and 3-D confinement.

The authors gratefully acknowledge assistance from Y.~Jiang, and financial support from NIST, NSF-PFC, AFOSR, and ARO with funding from the DARPA-OLE.


\begin{thebibliography}{29}
\bibitem[]{Footnote1}
\bibinfo{comment}{Shift cancellation occurs near 50$\%$ mean excitation for even large atomic sample inhomogeneity; however in the limit of small inhomogeneity, a given atom experiences the same shift in each atomic state and thus a shift cancellation on the transition.}
\end{thebibliography}
\end{document}